\documentclass[12pt,a4paper]{article}
\makeatletter
\def\eqnarray{%
\stepcounter{equation}%
\let\@currentlabel=\theequation
\global\@eqnswtrue
\global\@eqcnt\z@
\tabskip\@centering
\let\\=\@eqncr
$$\halign to \displaywidth\bgroup\@eqnsel\hskip\@centering
$\displaystyle\tabskip\z@{##}$&\global\@eqcnt\@ne
\hfil$\displaystyle{{}##{}}$\hfil
&\global\@eqcnt\tw@$\displaystyle\tabskip\z@{##}$\hfil
\tabskip\@centering&\llap{##}\tabskip\z@\cr}
\makeatother
\newcommand{\ket}[1]{{\vert{#1}\rangle}}

\begin{document}

\title{\sl A Rabi Oscillation in Four and Five Level Systems}
\author{
  Kazuyuki FUJII 
  \thanks{E-mail address : fujii@yokohama-cu.ac.jp },
  Kyoko HIGASHIDA 
  \thanks{E-mail address : s035577d@yokohama-cu.ac.jp },
  Ryosuke KATO 
  \thanks{E-mail address : s035559g@yokohama-cu.ac.jp }, 
  Yukako WADA 
  \thanks{E-mail address : s035588a@yokohama-cu.ac.jp }\\
  Department of Mathematical Sciences\\
  Yokohama City University\\
  Yokohama, 236--0027\\
  Japan
  }
\date{}
\maketitle
%
%
%
%
\begin{abstract}
  In this paper we consider a general model of an atom with n energy levels 
  interacting with n--1 external (laser) fields which is a natural extension 
  in the two and three level systems. 
  We exactly solve the Schr{\" o}dinger equation to obtain a Rabi oscillation 
  when n = 4 and 5, which will constitute a quantum logic gate of Quantum 
  Computation based on four and five level systems. 
\end{abstract}
%


%
%
%
%

\newpage
\section{Introduction}
The purpose of this paper is to construct a Rabi oscillation in 
four and five level systems which is a natural extension of two and three 
level systems, \cite{AE}, \cite{BR} or \cite{KF4}. 

That is, we consider a model of an atom with $n$ energy levels interacting 
with $n-1$ external (laser) fields and solve the Schr{\" o}dinger equation 
to obtain the Rabi oscillation. How to solve it can be reduced to 
the calculation of an exponential matrix, which is very clear from the 
mathematical view point. 
However, to calculate it in the general case seems almost impossible. 
For $n=4$ and $5$ we exactly calculate it by force to get the Rabi oscillation. 
The unitary matrix corresponding to the Rabi oscillation may play 
an important role in constructing the generalized Walsh--Hadamard matrix 
(transformation) or other useful matrices in four or five level systems. 

This is a succession of the preceeding paper \cite{FHKW} in which we treated 
the special case with coupling constants being all equal. 
Our motivation of the work is to construct Quantum Computation based on 
many energy levels (of laser--cooled atoms in a cavity). 
See also \cite{KF1}--\cite{KuF} for related work.

\section{General Theory}

We consider an atom with $n$ energy levels 
$
\{(\ket{k},E_{k})\ |\ 0 \leq k \leq n-1\}
$ 
which interacts with $n-1$ external fields. 
As for the external fields we use laser fields with frequencies equal to 
energy differences of the atom and set 
$
\Delta_{k}\equiv E_{k}-E_{0}\ \mbox{for}\ 1 \leq k \leq n-1. 
$ 
Moreover we assume 
\[
E_{1}-E_{0} > E_{2}-E_{1} > \cdots > E_{n-1}-E_{n-2}. 
\]

We subject the atom to $n-1$ laser fields having the frequencies $\omega_{k}$ 
equal to the energy differences $\Delta_{k}-\Delta_{k-1}=
E_{k}-E_{k-1}$. See the following picture : 

\vspace{-5mm}
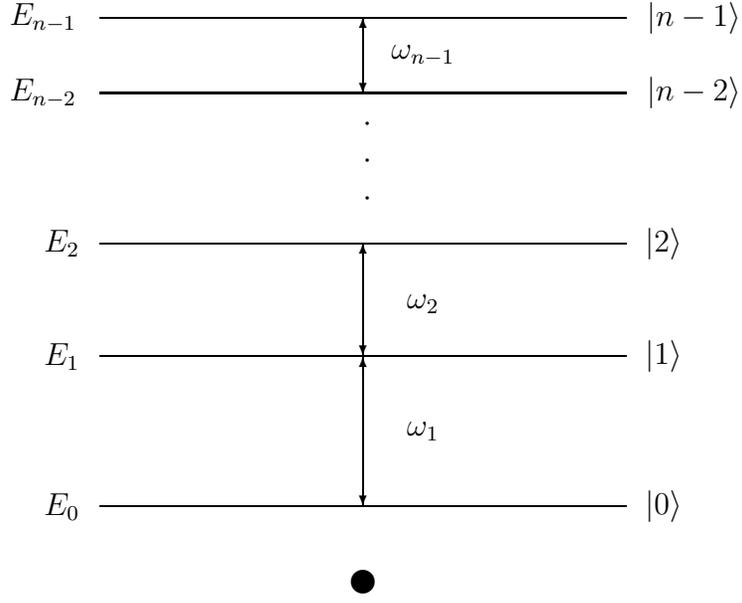
\begin{center}
\setlength{\unitlength}{1mm}   %
\begin{picture}(100,100)(0,0)
\put(0,80){\makebox(15,10)[c]{$E_{n-1}$}}
\put(15,85){\line(1,0){70}}
\put(85,80){\makebox(18,10)[c]{$|{n-1}\rangle$}}
\put(0,70){\makebox(15,10)[c]{$E_{n-2}$}}
\put(15,75){\line(1,0){70}}
\put(85,70){\makebox(18,10)[c]{$|{n-2}\rangle$}}
\put(5,50){\makebox(10,10)[c]{$E_2$}}
\put(15,55){\line(1,0){70}}
\put(85,50){\makebox(10,10)[c]{$|2\rangle$}}
\put(5,35){\makebox(10,10)[c]{$E_1$}}
\put(15,40){\line(1,0){70}}
\put(85,35){\makebox(10,10)[c]{$|1\rangle$}}
\put(5,15){\makebox(10,10)[c]{$E_0$}}
\put(15,20){\line(1,0){70}}
\put(85,15){\makebox(10,10)[c]{$|0\rangle$}}
\put(50,10){\circle*{3}}
\put(50,60){$\cdot$}
\put(50,65){$\cdot$}
\put(50,70){$\cdot$}
\put(50,30){\vector(0,1){10}}
\put(50,30){\vector(0,-1){10}}
\put(53,25){\makebox(10,10)[c]{$\omega_1$}}
\put(50,50){\vector(0,1){5}}
\put(50,50){\vector(0,-1){10}}
\put(53,42){\makebox(10,10)[c]{$\omega_2$}}
\put(50,80){\vector(0,1){5}}
\put(50,80){\vector(0,-1){5}}
\put(53,75){\makebox(10,10)[c]{$\omega_{n-1}$}}
\end{picture}
\end{center}

The Hamiltonian that we use is written as 
\begin{eqnarray}
\label{eq:hamiltonian-general}
&&H=                 \nonumber \\
&{}&                 \nonumber \\
&&
\left(
  \begin{array}{ccccccc}
   E_{0} & g_{1}\mbox{e}^{i(\omega_{1}t+\phi_{1})} &   &   &   &   &     \\
   g_{1}\mbox{e}^{-i(\omega_{1}t+\phi_{1})} & E_{1} &
   g_{2}\mbox{e}^{i(\omega_{2}t+\phi_{2})} &  &  &  &                    \\
     & g_{2}\mbox{e}^{-i(\omega_{2}t+\phi_{2})} & E_{2} &
    g_{3}\mbox{e}^{i(\omega_{3}t+\phi_{3})} &  &  &                      \\
     &  & g_{3}\mbox{e}^{-i(\omega_{3}t+\phi_{3})} & \cdot & \cdot &  &  \\
     &       &       & \cdot      & \cdot & \cdot  &                     \\
     &       &       &       & \cdot & 
       E_{n-2} & g_{n-1}\mbox{e}^{i(\omega_{n-1}t+\phi_{n-1})}           \\
    &  &  &  &  & g_{n-1}\mbox{e}^{-i(\omega_{n-1}t+\phi_{n-1})} & E_{n-1} 
  \end{array}
\right)                         \nonumber \\
&{}&                            \nonumber \\
&&=E_{0}{\bf 1}_{n}+            \nonumber \\
&&
\left(
  \begin{array}{ccccccc}
    0 & g_{1}\mbox{e}^{i(\omega_{1}t+\phi_{1})} &   &   &   &   &         \\
   g_{1}\mbox{e}^{-i(\omega_{1}t+\phi_{1})} & \Delta_{1} &
   g_{2}\mbox{e}^{i(\omega_{2}t+\phi_{2})} &  &  &  &                     \\
     & g_{2}\mbox{e}^{-i(\omega_{2}t+\phi_{2})} & \Delta_{2} &
    g_{3}\mbox{e}^{i(\omega_{3}t+\phi_{3})} &  &  &                       \\
     &  & g_{3}\mbox{e}^{-i(\omega_{3}t+\phi_{3})} & \cdot & \cdot &   &  \\
     &       &       & \cdot      & \cdot & \cdot &                       \\
     &       &       &       & \cdot & 
     \Delta_{n-2} & g_{n-1}\mbox{e}^{i(\omega_{n-1}t+\phi_{n-1})}         \\
   &  &  &  &  & g_{n-1}\mbox{e}^{-i(\omega_{n-1}t+\phi_{n-1})} & \Delta_{n-1} 
  \end{array}
\right)                         \nonumber \\
&{}&
\end{eqnarray}
where $g_{1},g_{2},\cdots,g_{n-1}$ are coupling constants and 
$\phi_{1},\phi_{2},\cdots,\phi_{n-1}$ some phases. Here we have assumed the 
so--called rotating wave approximation (RWA) from the beginning. 
We would like to decompose the Hamiltonian. If we set 
\begin{equation}
\label{eq:v(t)}
V=
\left(
  \begin{array}{ccccccc}
    1 &    &        &    &    &   &                                    \\
      & \mbox{e}^{i(\omega_{1}t+\phi_{1})} &  &   &   &  &             \\
      &    & \mbox{e}^{i(\omega_{1}t+\omega_{2}t+\phi_{1}+\phi_{2})} 
      &  &  &  &                                                       \\
      &    &           &  \cdot  &     &  &                            \\
      &    &           &         & \quad \quad \cdot   &    &          \\
      &    &  &  &  & \mbox{e}^{i\left(\sum_{k=1}^{n-2}\omega_{k}t+
      \sum_{k=1}^{n-2}\phi_{k}\right)} & \\
      &    &  &  &  &   &  \mbox{e}^{i\left(\sum_{k=1}^{n-1}\omega_{k}t+
      \sum_{k=1}^{n-1}\phi_{k}\right)} 
  \end{array}
\right),
\end{equation}
then it is easy to see 
\begin{equation}
H=V^{\dagger}
\left\{
E_{0}{\bf 1}_{n}+ 
\left(
  \begin{array}{ccccccc}
    0 & g_{1} &   &   &   &   &                          \\
   g_{1} & \Delta_{1} & g_{2} &   &   &   &              \\
     & g_{2} & \Delta_{2} & g_{3} &   &   &              \\
     &   & \cdot & \cdot & \cdot &   &                   \\
     &   &   & \cdot & \cdot & \cdot &                   \\
     &   &   &       & g_{n-2} & \Delta_{n-2} & g_{n-1}  \\
     &   &   &       &   & g_{n-1} & \Delta_{n-1} 
  \end{array}
\right)
\right\}
V. 
\end{equation}

To solve the Schr{\" o}dinger equation (we set $\hbar=1$ for simplicity) 
\begin{equation}
\label{eq:Schrodinger-equation}
i\frac{d}{dt}U=HU
\end{equation}
for $U(t)$ being unitary we set $\tilde{U}=VU$ ($\Leftrightarrow$ 
$U=V^{\dagger}\tilde{U}$), then it is not difficult to see 
\begin{eqnarray}
&&i\frac{d}{dt}\tilde{U}=     \nonumber \\
&{}&                          \nonumber \\
&&
\left\{
E_{0}{\bf 1}_{n}+ 
\left(
  \begin{array}{ccccccc}
    0 & g_{1} &   &   &   &   &                                       \\
   g_{1} & \Delta_{1}-\omega_{1} & g_{2} &   &   &   &                \\
     & g_{2} & \Delta_{2}-\omega_{1}-\omega_{2} & g_{3} &   &   &     \\
     &   & \cdot & \cdot & \cdot &    &                               \\
     &   &   & \cdot & \cdot & \cdot &                                \\
   &  &  &  & g_{n-2} & \Delta_{n-2}-\sum_{k=1}^{n-2}\omega_{k} & g_{n-1} \\
     &   &   &  &   & g_{n-1} & \Delta_{n-1}-\sum_{k=1}^{n-1}\omega_{k} 
  \end{array}
\right)
\right\}
\tilde{U}.                    \nonumber \\
&{}&
\end{eqnarray}

Now if we set the resonance conditions 
\begin{equation}
\label{eq:resonance-condition}
\Delta_{k}-\Delta_{k-1}=\omega_{k}\quad \mbox{for}\quad 1 \leq k \leq n-1,
\end{equation}
then the equation above reduces to a simple one 
\begin{equation}
i\frac{d}{dt}\tilde{U}=(E_{0}{\bf 1}_{n}+C)\tilde{U}
\end{equation}
where 
\begin{equation}
\label{eq:matrix-C}
C\equiv C(g_{1},g_{2},\cdots,g_{n-1})=
\left(
  \begin{array}{ccccccc}
    0 & g_{1} &   &   &   &   &             \\
    g_{1} & 0 & g_{2} &   &   &   &         \\
      & g_{2} & 0 & g_{3} &   &   &         \\
      &   & \cdot & \cdot & \cdot &  &      \\
      &   &   & \cdot & \cdot & \cdot &     \\
      &   &   &   & g_{n-2} & 0 & g_{n-1}   \\
      &   &   &   &   & g_{n-1} & 0
  \end{array}
\right).
\end{equation}
The solution is easily obtained to be 
$
\tilde{U}(t)=\mbox{e}^{-itE_{0}}\mbox{e}^{-itC}
$, 
so the solution we are looking for is 
\begin{equation}
U(t)=V^{\dagger}\tilde{U}(t)=\mbox{e}^{-itE_{0}}V^{\dagger}\mbox{e}^{-itC}. 
\end{equation}
Therefore we have only to calculate $\mbox{e}^{-itC}$, which is however 
very hard task. 

\vspace{5mm}
In the following sections let us list our calculations for $n=3,\ 4,\ 5$.

\section{Three Level Systems}

We only list the calculation in the case of three level systems. 
\begin{eqnarray}
\mbox{exp}(-itC)&=&\mbox{exp}
\left\{-it
\left(
  \begin{array}{ccc}
   0 & g_{1} & 0       \\
   g_{1}& 0  & g_{2}   \\
   0 & g_{2} & 0     
  \end{array}
\right)
\right\}                                \nonumber \\
&=&
\left(
  \begin{array}{ccc}
  \frac{g_{1}^{2}\mbox{cos}(\sqrt{g_{1}^{2}+g_{2}^{2}}t)+g_{2}^{2}}
  {g_{1}^{2}+g_{2}^{2}}& 
 -i\frac{g_{1}\mbox{sin}(\sqrt{g_{1}^{2}+g_{2}^{2}}t)}
  {\sqrt{g_{1}^{2}+g_{2}^{2}}}& 
  \frac{g_{1}g_{2}\mbox{cos}(\sqrt{g_{1}^{2}+g_{2}^{2}}t)-g_{1}g_{2}}
  {g_{1}^{2}+g_{2}^{2}}  \\
 -i\frac{g_{1}\mbox{sin}(\sqrt{g_{1}^{2}+g_{2}^{2}}t)}
  {\sqrt{g_{1}^{2}+g_{2}^{2}}}& \mbox{cos}(\sqrt{g_{1}^{2}+g_{2}^{2}}t)&
 -i\frac{g_{2}\mbox{sin}(\sqrt{g_{1}^{2}+g_{2}^{2}}t)}
  {\sqrt{g_{1}^{2}+g_{2}^{2}}} \\
  \frac{g_{1}g_{2}\mbox{cos}(\sqrt{g_{1}^{2}+g_{2}^{2}}t)-g_{1}g_{2}}
  {g_{1}^{2}+g_{2}^{2}}&
  -i\frac{g_{2}\mbox{sin}(\sqrt{g_{1}^{2}+g_{2}^{2}}t)}
  {\sqrt{g_{1}^{2}+g_{2}^{2}}}&
  \frac{g_{2}^{2}\mbox{cos}(\sqrt{g_{1}^{2}+g_{2}^{2}}t)+g_{1}^{2}}
  {g_{1}^{2}+g_{2}^{2}} 
  \end{array}
\right),
\end{eqnarray}
see Appendix in \cite{KF4} or \cite{AE}. 
As a result the solution we are looking for in the three level systems is 
\begin{eqnarray}
U(t)&=&\mbox{e}^{-itE_{0}}
\left(
  \begin{array}{ccc}
  1 & &                                                           \\
    & \mbox{e}^{-i(\omega_{1}t+\phi_{1})} &                       \\
    & & \mbox{e}^{-i(\omega_{1}t+\omega_{2}t+\phi_{1}+\phi_{2})}
  \end{array}
\right)\times             \nonumber \\
&&\left(
  \begin{array}{ccc}
  \frac{g_{1}^{2}\mbox{cos}(\sqrt{g_{1}^{2}+g_{2}^{2}}t)+g_{2}^{2}}
  {g_{1}^{2}+g_{2}^{2}}& 
 -i\frac{g_{1}\mbox{sin}(\sqrt{g_{1}^{2}+g_{2}^{2}}t)}
  {\sqrt{g_{1}^{2}+g_{2}^{2}}}& 
  \frac{g_{1}g_{2}\mbox{cos}(\sqrt{g_{1}^{2}+g_{2}^{2}}t)-g_{1}g_{2}}
  {g_{1}^{2}+g_{2}^{2}}  \\
 -i\frac{g_{1}\mbox{sin}(\sqrt{g_{1}^{2}+g_{2}^{2}}t)}
  {\sqrt{g_{1}^{2}+g_{2}^{2}}}& \mbox{cos}(\sqrt{g_{1}^{2}+g_{2}^{2}}t)&
 -i\frac{g_{2}\mbox{sin}(\sqrt{g_{1}^{2}+g_{2}^{2}}t)}
  {\sqrt{g_{1}^{2}+g_{2}^{2}}} \\
  \frac{g_{1}g_{2}\mbox{cos}(\sqrt{g_{1}^{2}+g_{2}^{2}}t)-g_{1}g_{2}}
  {g_{1}^{2}+g_{2}^{2}}&
  -i\frac{g_{2}\mbox{sin}(\sqrt{g_{1}^{2}+g_{2}^{2}}t)}
  {\sqrt{g_{1}^{2}+g_{2}^{2}}}&
  \frac{g_{2}^{2}\mbox{cos}(\sqrt{g_{1}^{2}+g_{2}^{2}}t)+g_{1}^{2}}
  {g_{1}^{2}+g_{2}^{2}} 
  \end{array}
\right).
\end{eqnarray}

\section{Four Level Systems}

We continue our calculation in the case of four level systems. 
We want to calculate 
\begin{equation}
\label{eq:target-matrix-4}
\mbox{exp}(-itC)=\mbox{exp}
\left\{-it
\left(
  \begin{array}{cccc}
   0 & g_{1} & 0 & 0           \\
   g_{1} & 0 & g_{2} & 0       \\
   0 & g_{2} & 0 & g_{3}       \\
   0 & 0 & g_{3} & 0
  \end{array}
\right)
\right\}
=
\left(
  \begin{array}{cccc}
    a_{11} & a_{12} & a_{13} & a_{14} \\
    a_{21} & a_{22} & a_{23} & a_{24} \\
    a_{31} & a_{32} & a_{33} & a_{34} \\
    a_{41} & a_{42} & a_{43} & a_{44} 
  \end{array}
\right)
\end{equation}
exactly. To make $C$ a diagonal form let us calculate the characteristic 
equation 
\begin{equation}
0=
|\lambda{\bf 1}_{4}-C|
=
\left|
  \begin{array}{cccc}
   \lambda & -g_{1} & 0 & 0            \\
   -g_{1} & \lambda & -g_{2} & 0       \\
   0 & -g_{2} & \lambda & -g_{3}       \\
   0 & 0 & -g_{3} & \lambda
  \end{array}
\right|            
=
\lambda^{4}-(g_{1}^{2}+g_{2}^{2}+g_{3}^{2})\lambda^{2}+g_{1}^{2}g_{3}^{2}.
\end{equation}
We can find solutions with good form (see \cite{KF5} as a modern derivation 
of the Ferrari formula). 
For that we set 
\[
A\equiv g_{2}^{2}+(g_{1}+g_{3})^{2},\quad B\equiv g_{2}^{2}+(g_{1}-g_{3})^{2},
\]
then all solutions are given by 
\begin{equation}
\lambda_{1}= \frac{\sqrt{A}+\sqrt{B}}{2}, \quad 
\lambda_{2}= \frac{\sqrt{A}-\sqrt{B}}{2}, \quad 
\lambda_{3}=-\frac{\sqrt{A}-\sqrt{B}}{2}, \quad 
\lambda_{4}=-\frac{\sqrt{A}+\sqrt{B}}{2}. 
\end{equation}
Let us rewrite these. It is easy to see 
\begin{equation}
\lambda_{1}= \frac{\sqrt{A}+\sqrt{B}}{2}\equiv \lambda, \quad 
\lambda_{2}= \frac{g_{1}g_{3}}{\lambda}, \quad 
\lambda_{3}=-\frac{g_{1}g_{3}}{\lambda}, \quad 
\lambda_{4}=-\lambda. 
\end{equation}

Next we must look for orthonormal eigenvectors $\ket{\lambda_{j}}$ ($j=1,\ 
2,\ 3,\ 4$) corresponding to eigenvalues above which are rather complicated. 
\begin{eqnarray}
\ket{\lambda_{1}}&=&
\left(
  \begin{array}{c}
    g_{1}g_{2}X                                     \\
    \lambda g_{2}X                                  \\
    (\lambda^{2}-g_{1}^{2})X                        \\
    \frac{g_{3}(\lambda^{2}-g_{1}^{2})}{\lambda}X
  \end{array}
\right),\quad 
\ket{\lambda_{2}}=
\left(
  \begin{array}{c}
    \lambda g_{2}Y                                  \\
    g_{2}g_{3}Y                                     \\
    \frac{g_{1}(g_{3}^{2}-\lambda^{2})}{\lambda}Y   \\
    (g_{3}^{2}-\lambda^{2})Y                        
  \end{array}
\right),                 \nonumber \\
\ket{\lambda_{3}}&=&
\left(
  \begin{array}{c}
   -\lambda g_{2}Y                                  \\
    g_{2}g_{3}Y                                     \\
   -\frac{g_{1}(g_{3}^{2}-\lambda^{2})}{\lambda}Y   \\
    (g_{3}^{2}-\lambda^{2})Y                        
  \end{array}
\right),\quad 
\ket{\lambda_{4}}=
\left(
  \begin{array}{c}
    g_{1}g_{2}X                                     \\
   -\lambda g_{2}X                                  \\
    (\lambda^{2}-g_{1}^{2})X                        \\
   -\frac{g_{3}(\lambda^{2}-g_{1}^{2})}{\lambda}X
  \end{array}
\right),
\end{eqnarray}
where $X$ and $Y$ are given by 
\begin{eqnarray}
X&=&\frac{1}{\sqrt{2\left\{
\lambda^{2}(-g_{1}^{2}+g_{2}^{2}+g_{3}^{2})+
g_{1}^{2}(g_{1}^{2}+g_{2}^{2}-g_{3}^{2})\right\}}},  \nonumber \\
Y&=&\frac{1}{\sqrt{2\left\{
g_{3}^{2}(-g_{1}^{2}+g_{2}^{2}+g_{3}^{2})+
\lambda^{2}(g_{1}^{2}+g_{2}^{2}-g_{3}^{2})
\right\}}}.                                          \nonumber 
\end{eqnarray}
Therefore we obtain the orthogonal matrix 
\begin{equation}
\label{eq:orthogonal-matrix-4}
W={\bf (}
\ket{\lambda_{1}},\ket{\lambda_{2}},\ket{\lambda_{3}},\ket{\lambda_{4}}
{\bf )}\ \in\ O(4),
\end{equation}
which makes $\mbox{exp}(-itC)$ an easy form to calculate 
\begin{equation}
\label{eq:diagonal-form-4}
\mbox{exp}(-itC)=W
\left(
  \begin{array}{cccc}
   \mbox{e}^{-it\lambda_{1}} &  &  &          \\
      & \mbox{e}^{-it\lambda_{2}} &  &        \\
      &   & \mbox{e}^{-it\lambda_{3}}  &      \\
      &   &  & \mbox{e}^{-it\lambda_{4}}
  \end{array}
\right)
W^{-1}.
\end{equation}
Let us calculate components of the matrix (\ref{eq:target-matrix-4}). 
From (\ref{eq:orthogonal-matrix-4}) and (\ref{eq:diagonal-form-4}) 
a long calculation leads to 
\begin{eqnarray}
\label{eq:components-4}
a_{11}&=&
2g_{2}^{2}\left\{g_{1}^{2}X^{2}\mbox{cos}(\lambda t)+
\lambda^{2}Y^{2}\mbox{cos}\left(\frac{g_{1}g_{3}}{\lambda}t\right)\right\}, 
                             \nonumber \\
a_{12}&=&a_{21}=
-2i\lambda g_{2}^{2}\left\{g_{1}X^{2}\mbox{sin}(\lambda t)+
g_{3}Y^{2}\mbox{sin}\left(\frac{g_{1}g_{3}}{\lambda}t\right)\right\}, 
                             \nonumber \\
a_{13}&=&a_{31}=
2g_{1}g_{2}\left\{(\lambda^{2}-g_{1}^{2})X^{2}\mbox{cos}(\lambda t)+
(g_{3}^{2}-\lambda^{2})Y^{2}\mbox{cos}\left(\frac{g_{1}g_{3}}{\lambda}t\right)
\right\}, 
                             \nonumber \\
a_{14}&=&a_{41}=
-2ig_{2}\left\{\frac{g_{1}g_{3}(\lambda^{2}-g_{1}^{2})}{\lambda}
X^{2}\mbox{sin}(\lambda t)+\lambda(g_{3}^{2}-\lambda^{2})
Y^{2}\mbox{sin}\left(\frac{g_{1}g_{3}}{\lambda}t\right)
\right\}, 
                             \nonumber \\
a_{22}&=&
2g_{2}^{2}\left\{\lambda^{2}X^{2}\mbox{cos}(\lambda t)+
g_{3}^{2}Y^{2}\mbox{cos}\left(\frac{g_{1}g_{3}}{\lambda}t\right)\right\}, 
                             \nonumber \\
a_{23}&=&a_{32}=
-2ig_{2}\left\{\lambda(\lambda^{2}-g_{1}^{2})X^{2}\mbox{sin}(\lambda t)+
\frac{g_{1}g_{3}(g_{3}^{2}-\lambda^{2})}{\lambda}
Y^{2}\mbox{sin}\left(\frac{g_{1}g_{3}}{\lambda}t\right)\right\}, 
                             \nonumber \\
a_{24}&=&a_{42}=
2g_{2}g_{3}\left\{(\lambda^{2}-g_{1}^{2})X^{2}\mbox{cos}(\lambda t)+
(g_{3}^{2}-\lambda^{2})Y^{2}\mbox{cos}\left(\frac{g_{1}g_{3}}{\lambda}t\right)
\right\}, 
                             \nonumber \\
a_{33}&=&
2\left\{(\lambda^{2}-g_{1}^{2})^{2}X^{2}\mbox{cos}(\lambda t)+
\frac{g_{1}^{2}(g_{3}^{2}-\lambda^{2})^{2}}{\lambda^{2}}
Y^{2}\mbox{cos}\left(\frac{g_{1}g_{3}}{\lambda}t\right)
\right\}, 
                             \nonumber \\
a_{34}&=&a_{43}=
-2i\frac{1}{\lambda}
\left\{g_{3}(\lambda^{2}-g_{1}^{2})^{2}X^{2}\mbox{sin}(\lambda t)+
g_{1}(g_{3}^{2}-\lambda^{2})^{2}
Y^{2}\mbox{sin}\left(\frac{g_{1}g_{3}}{\lambda}t\right)\right\},
                             \nonumber \\
a_{44}&=&
2\left\{\frac{g_{3}^{2}(\lambda^{2}-g_{1}^{2})^{2}}{\lambda^{2}}
X^{2}\mbox{cos}(\lambda t)+(g_{3}^{2}-\lambda^{2})^{2}
Y^{2}\mbox{cos}\left(\frac{g_{1}g_{3}}{\lambda}t\right)\right\}. 
\end{eqnarray}

\vspace{3mm}
As a result the solution we are looking for in the four level systems is 
\begin{eqnarray}
U(t)&=&\mbox{e}^{-itE_{0}}
\left(
  \begin{array}{cccc}
  1 & &                                                           \\
    & \mbox{e}^{-i(\omega_{1}t+\phi_{1})} &                       \\
    & & \mbox{e}^{-i(\omega_{1}t+\omega_{2}t+\phi_{1}+\phi_{2})}  \\
    & & & \mbox{e}^{-i(\omega_{1}t+\omega_{2}t+\omega_{3}t+
    \phi_{1}+\phi_{2}+\phi_{3})} 
  \end{array}
\right)\times        \nonumber \\
&&
\left(
  \begin{array}{cccc}
    a_{11} & a_{12} & a_{13} & a_{14} \\
    a_{21} & a_{22} & a_{23} & a_{24} \\
    a_{31} & a_{32} & a_{33} & a_{34} \\
    a_{41} & a_{42} & a_{43} & a_{44} 
  \end{array}
\right).
\end{eqnarray}
We obtained the explicit solution of (\ref{eq:Schrodinger-equation}) under 
the resonance condition (\ref{eq:resonance-condition}). This is the Rabi 
oscillation that we want, which is rather complicated.

\section{Five Level Systems}

We want to calculate 
\begin{equation}
\label{eq:target-matrix-5}
\mbox{exp}(-itC)=\mbox{exp}
\left\{-it
\left(
  \begin{array}{ccccc}
   0 & g_{1} & 0 & 0 & 0         \\
   g_{1} & 0 & g_{2} & 0 & 0     \\
   0 & g_{2} & 0 & g_{3} & 0     \\
   0 & 0 & g_{3} & 0 & g_{4}     \\
   0 & 0 & 0 & g_{4} & 0
  \end{array}
\right)
\right\}
=
\left(
  \begin{array}{ccccc}
    a_{11} & a_{12} & a_{13} & a_{14} & a_{15}  \\
    a_{21} & a_{22} & a_{23} & a_{24} & a_{25}  \\
    a_{31} & a_{32} & a_{33} & a_{34} & a_{35}  \\
    a_{41} & a_{42} & a_{43} & a_{44} & a_{45}  \\
    a_{51} & a_{52} & a_{53} & a_{54} & a_{55}
  \end{array}
\right)
\end{equation}
exactly. The characteristic equation is 
\begin{eqnarray}
0&=&
|\lambda{\bf 1}_{5}-C|
=
\left|
  \begin{array}{ccccc}
   \lambda & -g_{1} & 0 & 0 & 0           \\
   -g_{1} & \lambda & -g_{2} & 0 & 0      \\
   0 & -g_{2} & \lambda & -g_{3} & 0      \\
   0 & 0 & -g_{3} & \lambda & -g_{4}      \\
   0 & 0 & 0 & -g_{4} & \lambda 
  \end{array}
\right|          \nonumber \\
&=&
\lambda\left\{
\lambda^{4}-(g_{1}^{2}+g_{2}^{2}+g_{3}^{2}+g_{4}^{2})\lambda^{2}+
(g_{1}^{2}g_{3}^{2}+g_{1}^{2}g_{4}^{2}+g_{2}^{2}g_{4}^{2})
\right\}. 
\end{eqnarray}
All solutions are given by 
\begin{eqnarray}
\lambda_{1}&=& \frac{\sqrt{A+2\sqrt{B}}+\sqrt{A-2\sqrt{B}}}{2},\quad 
\lambda_{2}= \frac{\sqrt{A+2\sqrt{B}}-\sqrt{A-2\sqrt{B}}}{2},\quad 
\lambda_{3}=0,      \nonumber \\
\lambda_{4}&=&-\frac{\sqrt{A+2\sqrt{B}}-\sqrt{A-2\sqrt{B}}}{2},\quad 
\lambda_{5}=-\frac{\sqrt{A+2\sqrt{B}}+\sqrt{A-2\sqrt{B}}}{2}, 
\end{eqnarray}
where 
\[
A\equiv g_{1}^{2}+g_{2}^{2}+g_{3}^{2}+g_{4}^{2},\quad 
B\equiv g_{1}^{2}g_{3}^{2}+g_{1}^{2}g_{4}^{2}+g_{2}^{2}g_{4}^{2}.
\]
For 
\begin{equation}
\lambda_{1}=\frac{\sqrt{A+2\sqrt{B}}+\sqrt{A-2\sqrt{B}}}{2}\equiv \lambda, \
\lambda_{2}= \frac{\sqrt{B}}{\lambda}, \
\lambda_{3}=0, \ 
\lambda_{4}=-\lambda_{2}, \
\lambda_{5}=-\lambda 
\end{equation}
orthonormal eigenvectors $\ket{\lambda_{j}}$ ($j=1,\ 2,\ 3,\ 4,\ 5$) 
corresponding to the eigenvalues 
\begin{eqnarray}
\ket{\lambda_{1}}&=&
\left(
  \begin{array}{c}
    g_{1}g_{2}g_{3}X                                 \\
    \lambda g_{2}g_{3}X                              \\
    (\lambda^{2}-g_{1}^{2})g_{3}X                    \\
    \lambda(\lambda^{2}-g_{1}^{2}-g_{2}^{2})X        \\
    (\lambda^{2}-g_{1}^{2}-g_{2}^{2})g_{4}X
  \end{array}
\right),\
\ket{\lambda_{2}}=
\left(
  \begin{array}{c}
    \lambda g_{1}g_{2}g_{3}Y                                             \\
    \sqrt{B}g_{2}g_{3}Y                                                  \\
    \lambda g_{3}\left(\frac{B}{\lambda^{2}}-g_{1}^{2}\right)Y           \\
    \sqrt{B}\left(\frac{B}{\lambda^{2}}-g_{1}^{2}-g_{2}^{2}\right)Y      \\ 
    \lambda g_{4}\left(\frac{B}{\lambda^{2}}-g_{1}^{2}-g_{2}^{2}\right)Y  
  \end{array}
\right),\
\ket{\lambda_{3}}=
\left(
  \begin{array}{c}
       \frac{g_{2}g_{4}}{\sqrt{B}}  \\
       0                \\
      -\frac{g_{1}g_{4}}{\sqrt{B}}  \\
       0                \\
       \frac{g_{1}g_{3}}{\sqrt{B}}  
  \end{array}
\right),                     \nonumber \\
\ket{\lambda_{4}}&=&
\left(
  \begin{array}{c}
   -\lambda g_{1}g_{2}g_{3}Y                                             \\
    \sqrt{B}g_{2}g_{3}Y                                                  \\
   -\lambda g_{3}\left(\frac{B}{\lambda^{2}}-g_{1}^{2}\right)Y           \\
    \sqrt{B}\left(\frac{B}{\lambda^{2}}-g_{1}^{2}-g_{2}^{2}\right)Y      \\
   -\lambda g_{4}\left(\frac{B}{\lambda^{2}}-g_{1}^{2}-g_{2}^{2}\right)Y  
  \end{array}
\right),\
\ket{\lambda_{5}}=
\left(
  \begin{array}{c}
    g_{1}g_{2}g_{3}X                                 \\
   -\lambda g_{2}g_{3}X                              \\
    (\lambda^{2}-g_{1}^{2})g_{3}X                    \\
   -\lambda(\lambda^{2}-g_{1}^{2}-g_{2}^{2})X        \\
    (\lambda^{2}-g_{1}^{2}-g_{2}^{2})g_{4}X
  \end{array}
\right),
\end{eqnarray}
where $X$ and $Y$ are given by 
\begin{eqnarray}
X&=&\frac{1}{\sqrt{2\left\{(g_{3}^{2}+g_{4}^{2})A-2B\right\}\lambda^{2}+
2(g_{1}^{2}+g_{2}^{2}-g_{3}^{2}-g_{4}^{2})B}},  \nonumber \\
Y&=&\frac{1}{\sqrt{2B\left\{(g_{3}^{2}+g_{4}^{2})A-2B+
(g_{1}^{2}+g_{2}^{2}-g_{3}^{2}-g_{4}^{2})\lambda^{2}\right\}}}. 
\nonumber 
\end{eqnarray}
Therefore by the orthogonal matrix 
\begin{equation}
\label{eq:orthogonal-matrix-5}
W={\bf (}
\ket{\lambda_{1}},\ket{\lambda_{2}},\ket{\lambda_{3}},\ket{\lambda_{4}},
\ket{\lambda_{5}}
{\bf )}\ \in\ O(5),
\end{equation}
we have a diagonal form 
\begin{equation}
\label{eq:diagonal-form-5}
\mbox{exp}(-itC)=W
\left(
  \begin{array}{ccccc}
   \mbox{e}^{-it\lambda_{1}} &  &  &  &        \\
      & \mbox{e}^{-it\lambda_{2}} &  &  &      \\
      &  & \mbox{e}^{-it\lambda_{3}} &  &      \\
      &  &  & \mbox{e}^{-it\lambda_{4}} &      \\
      &  &  &  & \mbox{e}^{-it\lambda_{5}}
  \end{array}
\right)
W^{-1}.
\end{equation}
Let us calculate components of the matrix (\ref{eq:target-matrix-5}). 
A long calculation leads to 
\begin{eqnarray}
\label{eq:components-5}
a_{11}&=&\frac{g_{2}^{2}g_{4}^{2}}{B}+
2g_{1}^{2}g_{2}^{2}g_{3}^{2}\left\{X^{2}\mbox{cos}(\lambda t)+
\lambda^{2}Y^{2}\mbox{cos}\left(\frac{\sqrt{B}}{\lambda}t\right)\right\}, 
\nonumber \\
a_{12}&=&a_{21}=-2i\lambda g_{1}g_{2}^{2}g_{3}^{2}
\left\{X^{2}\mbox{sin}(\lambda t)+
\sqrt{B}Y^{2}\mbox{sin}\left(\frac{\sqrt{B}}{\lambda}t\right)\right\}, 
\nonumber \\
a_{13}&=&a_{31}=-\frac{g_{1}g_{2}g_{4}^{2}}{B}+2g_{1}g_{2}g_{3}^{2}
\left\{(\lambda^{2}-g_{1}^{2})X^{2}\mbox{cos}(\lambda t)+
(B-\lambda^{2} g_{1}^{2})Y^{2}\mbox{cos}\left(\frac{\sqrt{B}}{\lambda}t\right)
\right\},  \nonumber \\
a_{14}&=&a_{41}=-2i\lambda g_{1}g_{2}g_{3}
\left\{(\lambda^{2}-g_{1}^{2}-g_{2}^{2})X^{2}\mbox{sin}(\lambda t)+
\sqrt{B}\left(\frac{B}{\lambda^{2}}-g_{1}^{2}-g_{2}^{2}\right)
Y^{2}\mbox{sin}\left(\frac{\sqrt{B}}{\lambda}t\right)\right\},
\nonumber \\
a_{15}&=&a_{51}=g_{1}g_{2}g_{3}g_{4}\left\{\frac{1}{B}+
2(\lambda^{2}-g_{1}^{2}-g_{2}^{2})X^{2}\mbox{cos}(\lambda t)+
2(B-\lambda^{2}g_{1}^{2}-\lambda^{2}g_{2}^{2})
Y^{2}\mbox{cos}\left(\frac{\sqrt{B}}{\lambda}t\right)\right\},  
\nonumber \\
a_{22}&=&
2g_{2}^{2}g_{3}^{2}\left\{\lambda^{2}X^{2}\mbox{cos}(\lambda t)+
BY^{2}\mbox{cos}\left(\frac{\sqrt{B}}{\lambda}t\right)\right\},
\nonumber \\
a_{23}&=&a_{32}=-2i\lambda g_{2}g_{3}^{2}\left\{
(\lambda^{2}-g_{1}^{2})X^{2}\mbox{sin}(\lambda t)+
\sqrt{B}\left(\frac{B}{\lambda^{2}}-g_{1}^{2}\right)
Y^{2}\mbox{sin}\left(\frac{\sqrt{B}}{\lambda}t\right)\right\},
\nonumber \\
a_{24}&=&a_{42}=2g_{2}g_{3}\left\{
\lambda^{2}(\lambda^{2}-g_{1}^{2}-g_{2}^{2})X^{2}\mbox{cos}(\lambda t)+
B\left(\frac{B}{\lambda^{2}}-g_{1}^{2}-g_{2}^{2}\right)
Y^{2}\mbox{cos}\left(\frac{\sqrt{B}}{\lambda}t\right)\right\}, 
\nonumber \\
a_{25}&=&a_{52}=-2i\lambda g_{2}g_{3}g_{4}\left\{
(\lambda^{2}-g_{1}^{2}-g_{2}^{2})X^{2}\mbox{sin}(\lambda t)+
\sqrt{B}\left(\frac{B}{\lambda^{2}}-g_{1}^{2}-g_{2}^{2}\right)
Y^{2}\mbox{sin}\left(\frac{\sqrt{B}}{\lambda}t\right)\right\},
\nonumber \\
a_{33}&=&\frac{g_{1}^{2}g_{4}^{2}}{B}+2g_{3}^{2}\left\{
(\lambda^{2}-g_{1}^{2})^{2}X^{2}\mbox{cos}(\lambda t)+
\lambda^{2}\left(\frac{B}{\lambda^{2}}-g_{1}^{2}\right)^{2}
Y^{2}\mbox{cos}\left(\frac{\sqrt{B}}{\lambda}t\right)\right\}, 
\nonumber \\
a_{34}&=&a_{43}=-2i\lambda g_{3}\left\{(\lambda^{2}-g_{1}^{2})
(\lambda^{2}-g_{1}^{2}-g_{2}^{2})X^{2}\mbox{sin}(\lambda t)+
\sqrt{B}\left(\frac{B}{\lambda^{2}}-g_{1}^{2}\right)\times 
\right. \nonumber \\
&&\left.\qquad \qquad \qquad 
\left(\frac{B}{\lambda^{2}}-g_{1}^{2}-g_{2}^{2}\right)
Y^{2}\mbox{sin}\left(\frac{\sqrt{B}}{\lambda}t\right)\right\},
\nonumber \\
a_{35}&=&a_{53}=-\frac{g_{1}^{2}g_{3}g_{4}}{B}+2g_{3}g_{4}\left\{
(\lambda^{2}-g_{1}^{2})(\lambda^{2}-g_{1}^{2}-g_{2}^{2})
X^{2}\mbox{cos}(\lambda t)+
\lambda^{2}\left(\frac{B}{\lambda^{2}}-g_{1}^{2}\right)\times 
\right. \nonumber \\
&&\left.\qquad \qquad \qquad 
\left(\frac{B}{\lambda^{2}}-g_{1}^{2}-g_{2}^{2}\right)
Y^{2}\mbox{cos}\left(\frac{\sqrt{B}}{\lambda}t\right)\right\},
\nonumber \\
a_{44}&=&2\lambda^{2}(\lambda^{2}-g_{1}^{2}-g_{2}^{2})^{2}
X^{2}\mbox{cos}(\lambda t)+
2B\left(\frac{B}{\lambda^{2}}-g_{1}^{2}-g_{2}^{2}\right)^{2}
Y^{2}\mbox{cos}\left(\frac{\sqrt{B}}{\lambda}t\right),
\nonumber \\
a_{45}&=&a_{54}=-2i\lambda g_{4}\left\{(\lambda^{2}-g_{1}^{2}-g_{2}^{2})^{2}
X^{2}\mbox{sin}(\lambda t)+
\sqrt{B}\left(\frac{B}{\lambda^{2}}-g_{1}^{2}-g_{2}^{2}\right)^{2}
Y^{2}\mbox{sin}\left(\frac{\sqrt{B}}{\lambda}t\right)\right\},
\nonumber \\
a_{55}&=&\frac{g_{1}^{2}g_{3}^{2}}{B}+2g_{4}^{2}\left\{
(\lambda^{2}-g_{1}^{2}-g_{2}^{2})^{2}X^{2}\mbox{cos}(\lambda t)+
\lambda^{2}\left(\frac{B}{\lambda^{2}}-g_{1}^{2}-g_{2}^{2}\right)^{2}
Y^{2}\mbox{cos}\left(\frac{\sqrt{B}}{\lambda}t\right)\right\}. 
\end{eqnarray}

\vspace{5mm}
As a result the solution we are looking for in the five level systems is 
\begin{eqnarray}
U(t)&=&\mbox{e}^{-itE_{0}}
\left(
  \begin{array}{ccccc}
  1 & &                                                                   \\
    & \mbox{e}^{-i(\omega_{1}t+\phi_{1})} &                               \\
    & & \mbox{e}^{-i(\omega_{1}t+\omega_{2}t+\phi_{1}+\phi_{2})}          \\
    & & & \mbox{e}^{-i(\sum_{k=1}^{3}\omega_{k}t+\sum_{k=1}^{3}\phi_{k})} \\
    & & & & 
    \mbox{e}^{-i(\sum_{k=1}^{4}\omega_{k}t+\sum_{k=1}^{4}\phi_{k})}
  \end{array}
\right)\times        \nonumber \\
&&
\left(
  \begin{array}{ccccc}
    a_{11} & a_{12} & a_{13} & a_{14} & a_{15}  \\
    a_{21} & a_{22} & a_{23} & a_{24} & a_{25}  \\
    a_{31} & a_{32} & a_{33} & a_{34} & a_{35}  \\
    a_{41} & a_{42} & a_{43} & a_{44} & a_{45}  \\
    a_{51} & a_{52} & a_{53} & a_{54} & a_{55}
  \end{array}
\right).
\end{eqnarray}
We obtained the explicit solution of (\ref{eq:Schrodinger-equation}) under 
the resonance condition (\ref{eq:resonance-condition}). This is the Rabi 
oscillation that we want, which is quite complicated.

\section{Discussion}

We have calculated $\mbox{e}^{-itC}$ with (\ref{eq:matrix-C}) 
for $n=4$ and $5$ which is a generalization of the case $n=3$ in \cite{AE} or 
\cite{KF4}. However, to calculate the general case is a very hard problem. 
One reason is as follows. In the general case the characteristic equation 
$|\lambda{\bf 1}-C|$ becomes a general algebraic equation with degee n. 
By the Galois theory it is impossible to solve it in an algebraic manner. 
That is, we have no algebraic method to write down the solutions $\lambda_{k}$. 
\par \noindent
For the special case $g_{1}=g_{2}=\cdots=g_{n-1}\equiv g$ we can calculate 
it easily, see \cite{FHKW}. 

We also note that problems of this type occur elsewhere in Mathematics and 
Physics, see for example \cite{CDEL}. Thus our results may have broader 
applicability.

\vspace{10mm}
\noindent{\em Acknowledgment.}\\
We wishes to thank Tatsuo Suzuki for his helpful comments and suggestions.


\end{document}